\documentclass[sigconf]{acmart}

\AtBeginDocument{%
  \providecommand\BibTeX{{%
    \normalfont B\kern-0.5em{\scshape i\kern-0.25em b}\kern-0.8em\TeX}}}

\copyrightyear{2025}
\acmYear{2025}
\setcopyright{acmlicensed}\acmConference[KDD '25]{Proceedings of the 31st ACM SIGKDD Conference on Knowledge Discovery and Data Mining V.1}{August 3--7, 2025}{Toronto, ON, Canada}
\acmBooktitle{Proceedings of the 31st ACM SIGKDD Conference on Knowledge Discovery and Data Mining V.1 (KDD '25), August 3--7, 2025, Toronto, ON, Canada}
\acmDOI{10.1145/3690624.3709431}
\acmISBN{979-8-4007-1245-6/25/08}

\usepackage{acronym} %
\usepackage{amsmath}
\usepackage{balance} %
\usepackage{bm} %
\usepackage{hyperref} %
\usepackage{optidef} %
\usepackage{subfig} %

\newcommand{\sat}{ForTune}%
\newcommand{\match}{Matching}%
\newcommand{\criteo}{\href{https://www.criteo.com/}{Criteo}}%
\newcommand{\spotify}{\href{https://open.spotify.com/}{Spotify}}%
\newcommand{\spotinc}{Spotify, Inc.}%
\newcommand{\cf}[1]{\ensuremath{\mathtt{f}_{#1}}}%
\newcommand{\visit}{\ensuremath{v}}%
\newcommand{\tv}{\ensuremath{\visit_{\text{test}}}}%
\newcommand{\etv}{\ensuremath{\hat{\visit}_{\text{test}}}}%
\DeclareRobustCommand{\bbone}{\text{\usefont{U}{bbold}{m}{n}1}}

\acrodef{iws}[IWS]{Inverse Probability Weighting}%
\acrodef{psm}[PSM]{Propensity Score Matching} %
\acrodef{nnm}[NNM]{Nearest Neighbor Matching} %
\acrodef{ope}[OPE]{Off-Policy Evaluation} %
\acrodef{rl}[RL]{Reinforcement Learning} %
\acrodef{lci}[LCI]{Long-Term Causal Inference}%

\title{ForTune: Running Offline Scenarios to Estimate Impact on Business Metrics}%

\author{Georges Dupret}
\orcid{0000-0001-5744-2545}
\affiliation{%
  \institution{\spotinc}
  \city{New York}
  \state{NY}
  \country{USA}}
\email{gdupret@spotify.com}

\author{Konstantin Sozinov}
\orcid{0009-0005-4593-0994}
\affiliation{%
  \institution{\spotinc}
  \city{New York}
  \state{NY}
  \country{USA}}

\author{Carmen Barcena Gonzalez}
\orcid{0009-0005-1418-014X}
\affiliation{%
  \institution{\spotinc}
  \city{London}
\country{UK}
}

\author{Ziggy Zacks}
\orcid{0009-0007-8607-4121}
\affiliation{%
  \institution{\spotinc}
  \city{New York}
  \state{NY}
  \country{USA}}

\author{Amber Yuan}
\orcid{0000-0002-7945-7665}
\affiliation{%
  \institution{\spotinc}
  \city{New York}
  \state{NY}
  \country{USA}}

\author{Ben Carterette}
\orcid{0000-0001-9538-047X}
\affiliation{%
  \institution{\spotinc}
  \city{New York}
  \state{NY}
  \country{USA}}
\email{benc@spotify.com}

\author{Manuel Mai}
\orcid{0009-0000-6739-0486}
\affiliation{%
  \institution{\spotinc}
  \city{Berlin}
  \country{Germany}}

\author{Andrey Gatash}
\orcid{0009-0001-2024-3996}
\affiliation{%
  \institution{\spotinc}
  \city{New York}
  \state{NY}
  \country{USA}}

\author{Leo Lien}
\orcid{0009-0003-8062-0524}
\affiliation{%
  \institution{\spotinc}
  \city{New York}
  \state{NY}
  \country{USA}}

\author{Shubham Bansal}
\orcid{0009-0000-2410-6246}
\affiliation{%
  \institution{\spotinc}
  \city{New York}
  \state{NY}
  \country{USA}}

\author{Roberto Sanchis-Ojeda}
\orcid{0000-0002-6193-972X}
\affiliation{%
  \institution{\spotinc}
  \city{Madrid}
  \country{Spain}}

\author{Mounia Lalmas-Roelleke}
\orcid{0000-0002-3531-3096}
\affiliation{%
  \institution{\spotinc}
  \city{London}
  \country{UK}}

\ccsdesc[500]{Applied computing}
\ccsdesc[500]{Applied computing~Decision analysis}

\keywords{Scenario Analysis, Sensitivity Analysis, A/B-testing
  Prediction, A/B-testing Setup }

\date{\today}%

\begin{document}

\begin{abstract}

Making ideal decisions as a product leader in a web-facing company is incredibly challenging. Beyond navigating the ambiguity of customer satisfaction and achieving business goals, leaders must also ensure their products and services remain relevant, desirable, and profitable. Data and experimentation are crucial for testing product hypotheses and informing decisions. Online controlled experiments, such as A/B testing, can provide highly reliable data to support decisions. However, these experiments can be time-consuming and costly, particularly when assessing impacts on key business metrics like retention or long-term value.

Offline experimentation allows for rapid iteration and testing but often lacks the same level of confidence and clarity regarding business metrics impact. To address this, we introduce a novel, lightweight, and flexible approach called {\em scenario analysis}. This method aims to support product leaders’ decisions by using user data and estimates of business metrics. While it cannot fully replace online experiments, it offers valuable insights into trade-offs involved in growth or consumption shifts, estimates trends in long-term outcomes like retention, and can generate hypotheses about relationships between metrics at scale.
  
We implemented scenario analysis in a tool named \sat. We conducted experiments with this tool using a publicly available dataset and reported the results of experiments carried out by \spotify, a large audio streaming service, using \sat\ in production. In both cases, the tool reasonably predicted the outcomes of controlled experiments, provided that features were carefully chosen. We demonstrate how this method was used to make strategic decisions regarding the impact of prioritizing one type of content over another at \spotify.
  
\end{abstract}

\maketitle

\section{Introduction}
\label{sec:intro}

Product leaders in web-facing companies continually face challenging decisions.  Their decisions impact customer satisfaction in unpredictable ways. Besides navigating the ambiguity of customer satisfaction and achieving business goals, leaders must also ensure that  products and services to remain relevant, desirable, and profitable. However, it is not always clear how a decision today will contribute future success.

To make well-informed informed decisions, companies often conduct numerous online experiments, with some running over 200 concurrent experiments~\cite{kohavi2013online}. However, running large number of experiments introduces significant complications.

From an infrastructural perspective, exposing users to multiple experiments simultaneously complicates evaluation, as noise and interaction effects can bias results. Additionally, controlled experiments can test only a limited number of hypotheses, requiring teams to choose configurations wisely. Furthermore, not all hypotheses can be tested online due to technical constraints or pre-allocated traffic to other experiments.

From an outcomes and analysis perspective, online experiments might not provide a holistic picture of how a new product or service affects users. Long-term outcomes such as user satisfaction, retention and financial metrics such as revenue and gross profits are difficult to measure in real-time, especially within the confines of a short experiment.

Given these challenges, an offline method that tests hypotheses and overcomes some limitations of online experimentation would be valuable. Specifically, a method that allows testing many configurations and projects the impact on long-term key metrics.

The methodology behind \sat, the tool we introduce in this paper, is designed to gain insight into what the results of a controlled experiment might look like before deploying one. To that aim, we rely on a ``scenario'' that describes, in general terms, how we think some key business metrics will be affected by the treatment. It can be understood as a kind of sensitivity analysis of the business metrics with respect to a set of control variables, but without the need of an explicit model. It is not meant to replace controlled experiments, but rather to help prioritize between several experiment candidates. It can also be used to make an educated guess when it is not possible to set up a controlled experiment.

Few flexible and lightweight methods are available to carry out such predictions. Developing a predictive model is one possibility, but it requires deep system  knowledge and is typically complex and expensive in terms of computation and human effort. In contrast, the method we present is simple, flexible, and easy to implement.

For example, imagine we expect a 1\% increase in podcast consumption in a online audio app, such as \spotify, following an algorithmic or interface modification. Should we release this new version? Does an increase in podcast consumption lead to increased profit? What would be the impact on user retention? Would there be a substitution effect, leading to decreased music consumption? The only definitive way to answer these questions is to run a controlled experiment, which is time consuming and expensive, especially when evaluating long term metrics. 

Our proposed method predicts the impact of such a controlled
experiment before deploying changes and without needing to develop and
train a predictive model. The core idea is to collect past consumption
data and re-weight observations to match the expected changes. For
instance, returning to our example, we would expect more users
characterized by a high podcast consumption after deploying the app
change. If those users exhibit a higher retention rate, the test
branch of the experiment would also show a higher retention rate.

We successfully verified this idea on a publicly available dataset, a large controlled experiment from an advertising company, \criteo, that provides online display advertisements. We also verified the predictions using several controlled experiments with proprietary data from \spotify. \sat\ is both expressive and flexible, leading to its adoption at various levels within \spotify\ to inform strategic decision-making.

Section~\ref{sec:algo} presents the method. In Section~\ref{sec:algo:intuition} we use an example to illustrate it. We then extend the method to more general scenarios in Section~\ref{sec:algo:constraints} and summarize it in Section~\ref{sec:algo:formal}. We discuss caveat and limitations in Section~\ref{sec:algo:limitations} and we review related works in Section~\ref{sec:related}. Finally, we present experiments in Section~\ref{sec:experiments}: those conducted on the \criteo\ dataset in Section~\ref{sec:criteo}, and on \spotify\ proprietary data in Section~\ref{sec:spotify}.

\section{The \sat\ Algorithm}
\label{sec:algo}

We describe our proposed approach, \emph{\sat}. We begin with a simple example to build intuition and then generalize the solution to address more complex problems.

\subsection{The Intuition behind the Algorithm}
\label{sec:algo:intuition}

Suppose you own a shoe store and launch a campaign to double the number of male customers. How will this impact the average sale price? A simple estimate can be obtained by re-weighting past male customers twice as much as female customers (or resampling accordingly) and then computing the weighted average of sale prices. An example is provided in Table~\ref{tab:shoeshop}.

\begin{table}[h]
  \centering
  \begin{tabular}{rrrrrr}
    gender & age & shoe size & marital status & price & weight \\
    \hline
    F & 97 & 34 & married & 180 & 1 \\
    F & 85 & 53 & single & 150 & 1 \\
    M & 80 & 47 & single & 390 & 2 \\
    M & 45 & 49 & married & 180 & 2 \\
    M & 54 & 50 & single & 300 & 2 \\
    M & 79 & 54 & single & 340 & 2 \\
    F & 69 & 39 & married & 250 & 1
  \end{tabular}
  \caption{Hypothetical shoe store. Before the campaign, the average shoe sale is the average of the price column, i.e. \$256. After the campaign we expect twice as many male customers as before. We do not actually know the exact characteristics of these new customers, so a safe bet is to assume that they will be similar to the current male customers. Therefore, we replicate all male rows in the table. The estimated average sale price after the campaign is estimated as the average of the price column weighted by the weight column, i.e. \$273.}
  \label{tab:shoeshop}
\end{table}

This estimated average sale price will be accurate as long as the original male customers are representative of the new customers. However, if the campaign specifically targets males under 50 years old, and only one current customer fits this demographic, our prediction will rely exclusively on him, making it unreliable. Generally, we need enough samples to ensure our estimator controls the variance.

If we have more information about the new customers beyond gender, we can refine our predictions. For instance, if we target single male customers, a quick look at Table~\ref{tab:shoeshop} reveals that they spend more, suggesting the average sale price should be higher after the campaign. A similar conclusion applies if we target older male customers, such as those around 65 years of age.

In the next section, we formalize these ideas and propose a method to derive a set of weights that incorporate assumptions about the campaign's impact, while making minimal assumptions about customer characteristics.

\subsection{Building More Complex Scenarios}
\label{sec:algo:constraints}

In the previous section, we discussed a simple case where the only constraint on the population was the gender mix. Often, constraints apply to a continuous variable (like age) instead of a binary one. Additionally, there may be scenarios where multiple variables need to be constrained.

Let $X$ be a $N \times D$ feature matrix, and $y$ an $N$ dimensional vector representing the metric of interest. In Table~\ref{tab:shoeshop}, the features include gender, age, shoe size and marital status. The vector~$y$ represents the business metric we want to predict, which in this case is the price. The dimensions are $N=7$ and $D=4$.

The objective is to identify a $N$~dimensional vector $\bf{\omega}$ of weights that incorporate the expected change. In the shoe store example, this means doubling the proportion of male customers, i.e. denoting by $X_n$ the $n^{th}$~observation in $X$ and $\bbone_M(X_n)$~the indicator function for males:
\begin{align}
  \label{eq:males}
  \frac 1N \sum_{n=1}^N \omega_n \bbone_M(X_{n}) &= \frac 2N \sum_{n=1}^N \bbone_M(X_{n})
\end{align}

This constraint requires that the weights~$\bm{\omega}$ adjust the proportion of males in~$X$ to be twice the original proportion, i.e twice $\frac 1N \sum_{n=1}^N \bbone_M(X_{n})$. The solution to this constraint is not unique; many different vectors $\bm{\omega}$ can satisfy it.

If the campaign also targets older males, an additional constraint must be enforced. For example, to ensure that the average age of male customers after the campaign is 65 years old, the following constraint would be added:
\begin{align}
  \label{eq:age65}
  \frac 1N \sum_{n=1}^N \omega_n \bbone_M(X_{n}) X_{n,\textnormal{age}} &= 65
\end{align}

Similarly, we could opt for a more relaxed constraint by only requiring that the average male age be greater than 65 years old:
\begin{align}
  \label{eq:age65b}
  \frac 1N \sum_{n=1}^N \omega_n \bbone_M(X_{n}) X_{n,\textnormal{age}} &\geq 65
\end{align}

There are still many vectors of weights that satisfy the constraints in Equation~\eqref{eq:males} and Equation~\eqref{eq:age65} or~\eqref{eq:age65b}. Solutions that give large weights to a small number of observations are undesirable because they would rely excessively on a subset of the data, leading to high variance. Instead, it is  reasonable to look for a solution that distributes importance more evenly across all observations, while remaining compatible with the constraints. This approach suggests maximizing the entropy of the weights while satisfying the constraints. The optimization problem can be formulated as follows:

\begin{argmaxi!}
{\bm{\omega}}{\mathcal{H}(\bm{\omega})}
{\label{form:sat}}{}
\addConstraint{\frac 1N \sum_{n=1}^N \omega_n \bbone_M(X_{n})}{=\frac 2N \sum_{n=1}^N \bbone_M(X_{n})}
\addConstraint{\frac 1N \sum_{n=1}^N \omega_n \bbone_M(X_{n}) X_{n,\textnormal{age}}}{=65}
\addConstraint{\omega_n}{\geq 0}{n \in \{1\dots N\} \label{eq:wpos}}
\addConstraint{\sum_n^N \omega_n}{=1}{n \in \{1\dots N\} \label{eq:sumone}}
\end{argmaxi!}

We added two natural constraints: the weights must be positive, and they must sum to one, as shown in in Equations~\eqref{eq:wpos} and \eqref{eq:sumone}.

If no constraints are imposed beyond these natural ones, the solution to this problem is to assign equal weights to each observation, i.e. $\omega = \frac 1N$. This aligns with the fact that, prior to the campaign, the target metrics are estimated as the sample average.

\subsection{The Algorithm}
\label{sec:algo:formal}

We now formalize the algorithm we built an intuition for in the previous two sections. 

The \sat\ Algorithm involves solving the following convex optimization problem:\footnote{We require the problem to be convex for convenience. We could generalize it to any type of constraints but we have not encountered in practice a case where this is useful.}
\begin{argmaxi}
{\bm{\omega}}{\mathcal{H}(\bm{\omega})}
{\label{form:genalgo}}{}
\addConstraint{g_i(x)}{\leq 0}{\quad i=1,\dots,m}
\addConstraint{h_j(x)}{=0}{\quad j=1,\dots,p}
\addConstraint{\omega_n}{\geq 0}{\quad n \in \{1\dots N\}}
\addConstraint{\sum_n^N \omega_n}{=1}{\quad n \in \{1\dots N\}}
\end{argmaxi}
Here, the $h_i$ are $p \geq 0$ affine functions and $g_j\leq 0$ are $m\geq 0$ convex inequalities. The last two constraints are the natural constraints. The functions~$g_i$ and $h_j$ define the scenario associated with the problem. The problem of finding the right set of weights can be solved using an appropriate convex solver, and the solution is unique.

Once the weights~$\bm{\omega}$ are known, the metric of interest~$t$ is estimated as
\begin{eqnarray}
  \label{eq:estimate}
  \hat{t} &= \sum_n^N \omega_n t_n
\end{eqnarray}

This algorithm provides a point estimate of the prediction under the scenario hypotheses. To obtain a distribution of point estimates and measure of uncertainty, the dataset can be divided into~$B$ distinct random subsets, and the point estimate can be independently calculated for each of subset. Another approach is bootstrapping~\cite{efron2000bootstrap}, where the dataset is repeatedly sampled with replacement to create~$B$ subsets. The experiments in Section~\ref{sec:experiments} will highlight the need to estimate uncertainty.

\subsection{Limitations}
\label{sec:algo:limitations}

\begin{figure}[t]
  \centering
  \caption{Box plots of the Resampling Weights for the \criteo\ dataset. The weights have been multiplied by the number of observations so a weight of 1 means that the observations has the same importance in the control and treatment branches. A weight of 5 means that the corresponding observations is five times more influential in the test branch than before resampling. We set the constraints on features~\cf{1}, \cf{4}, \cf{7} and \cf{10} to be same multiples of the corresponding averages. The multiples are reported on the y axis. The further from the original means (for which the multiple is 1.0), the larger the weights' spread.%
  }
  \includegraphics[width=.4\textwidth]{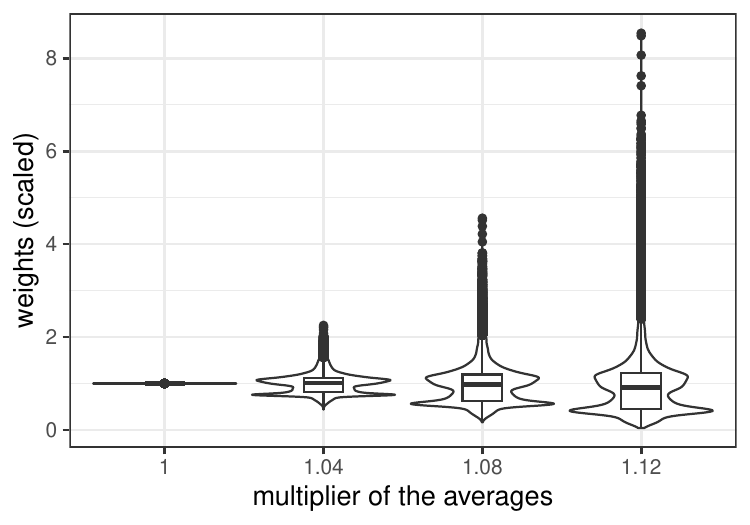}
  \label{fig:criteo:weights}
\end{figure}

Not all constraints are feasible or lead to realistic outcomes. It is important to remember that predictions are obtained by giving more importance to users who help satisfying the constraints. If no user can help, then the constraint is unrealizable. Returning to the example above, if the constraint required that the average male age be 100 years old instead of 65 (see Equation~\eqref{eq:age65}), no combination of weights would satisfy the convex problem. Less obvious sets of constraints might also be unrealizable, but in practice they are easy to detect because the quadratic solver will fail to return a solution.

A more subtle case arises when conditions are realizable, but the
solution assigns large weights to a limited number of observations,
potentially leading to a high variability. In the previous example,
this would occur if only a few users were aged 65 or older. In such
cases, the solver will give large weights to these few observations,
resulting in an estimate based on a small number of data points. This
problem can be diagnosed by examining the weights distribution to
identify extreme values. Figure~\ref{fig:criteo:weights} show the
effect of applying constraints set to 104, 108 and 112\% of the
original feature averages. More details are provided in
Section~\ref{sec:experiments} where we discuss the \criteo\ dataset,
but it should be clear that the further the constraints are from the
control data averages, the more likely we are to find outlier weights.

It is also important to remember that the~\sat\ is useful for
predicting averages, not absolute values. For instance, in the shoe
shop example from Section~\ref{sec:algo:intuition}, we predicted the
average sell price, not the total sales revenue. This is because the
campaign likely increased the number of customers by an amount that we
ignore. Not accounting for changes in the user base can lead to
apparent paradoxes. If the campaign targeted women instead of men, who
in this particular example buy cheaper shoes, the average price would
decrease even though total sales might increase due to the additional
customers.

As in observational studies~\cite{hernan2010causal}, obtaining an unbiased estimate of the effect requires identifying confounding covariates. This typically relies on domain knowledge, although automated methods have been developed to help identify causal graphs~\cite{scutari2010bnlearn}. At \spotify\ we built knowledge by comparing \sat\ predictions with the results of past controlled experiments. When predictions were off, we investigated the reasons and identified missing variables. An example is provided in Section~\ref{sec:spotify}.

\section{Related Works}
\label{sec:related}

This work began with an investigation into Sensitivity Analysis, which is typically defined as studying the relationship between model output uncertainty and each input variable. However, in practice it encompasses much more. As Iooss \& Lemaître~\cite{iooss2015review} note,  one application aligns with \sat: ``mapping the output behavior as a function of the inputs, focusing on a specific domain of inputs if necessary.''

Sensitivity Analysis can be divided into local and global methods~\cite{saltelli2009sensitivity}. Local Sensitivity Analysis focuses on the impact of small input perturbations around a nominal value of the features (typically the mean) on the model output. When the model is sufficiently simple, Taylor series expansions can approximate the model, allowing for an analytical differential sensitivity index to be derived~\cite{xu2008uncertainty}. Global sensitivity, on the other hand,  examines the model's overall response (averages over variations of all features) by exploring a finite region of the input domain. While early global sensitivity analyses technique assumed feature independence, newer approaches do not~\cite{xu2008uncertainty,li2010global}.

\sat\ can be viewed as a Sensitivity Analysis method because it
examines how output depends on input variations. However, there are
fundamental methodological differences and differences in
applicability. Sensitivity Analysis requires a model of the
relationship between inputs and outputs, which can incorporate
external knowledge (e.g., Bayesian models that explicitly define
relationships between features). In contrast, \sat\ only relies only
on resampling / re-weighting the data and implicitly accounts for
features correlations. While accurate feature selection is crucial for
\sat, this is also true for designing analytical models.

Another fundamental methodological difference between Sensitivity
Analysis and \sat\ is how the scenarios are defined. While the latter
typically requires only a set of global constraints on feature
averages to be set, the predictive models used in Sensitivity Analysis
requires all the individual features to be specified before
generating a prediction. To see why this is a hard problem, imagine
that \spotify\ is interested in estimating the impact on podcast
consumption of an increase of 5\% in music consumption. Do we increase
uniformly music consumption by 5\% before running the predictive
model? That is probably not realistic. Other features used by the
predictive model probably need adjustment because they are not
independent from music consumption. How is their values to be decided?

While \sat\ and Sensitivity Analysis share a similar goal, other methods share a similar approach. Weighting observations to query the data is common in statistics. Examples include stratified sampling~\cite{botev2017variance} and the Horvitz–Thompson estimator~\cite{horvitz1952generalization}. In election polling~\cite{weisberg1996introduction}, data from a small sample is used to predict larger population behavior. To ensure accuracy, cohorts within the sample are weighted to represent the broader population, considering factors such as race, age, gender, education, and geographical location. These factors play a significant role in voting behavior, and the demographics of the sample group should match the demographics of the voting population. The success of Stratified Sampling and \sat\ depends heavily on selecting the right set of features to match the original data to the target population.

\ac{iws} and \sat\ both aim to estimate quantities related to a target population different from the one the data was collected from. The key difference is that \ac{iws} assumes a sample of the target dataset is available, while \sat\ relies on a limited set of global constraints to characterize the target population. \ac{psm}~\cite{rosenbaum1983central} is another technique for estimating the effect of an intervention by accounting for covariates that predict receiving the treatment. While we use this method in Section~\ref{sec:criteo:match} to compare it with \sat\, unlike \acs{psm}, \sat\ does not require a treatment set. The process of matching is improved upon using Entropy maximization in~\cite{hainmueller2012entropy,lin2023balancing}.

Both \ac{iws} and \ac{psm} are used in {\em counterfactual analysis}, a crucial research area for hypothesis testing, offline evaluation, and learning in multi-armed bandits and \ac{rl} exploration policies~\cite{saito21}. {\em \ac{ope}}, a large subset of counterfactual analysis, estimates outcomes from deploying a target policy to a population from which data has already been collected. \acs{ope} uses existing data collected based on a logging policy (typically randomization) and re-weights it with propensities given by the target policy to estimate what would happen to quantities of interest if the target policy were to be deployed.  Research in \acs{ope} includes applications in slate recommendation~\cite{swaminathan17}, sequential search~\cite{miao22}), reducing the variance of \acs{ope}~\cite{dudik11}, leveraging historic data~\cite{agarwal17}, minimally-invasive randomized interventions~\cite{joachims16}, long-term off-policy estimation~\cite{saito24}, and interdependent reward models~\cite{mcinerney20}.

Approaches to long-term causal inference (\acs{lci}) are also relevant to our aims. \acs{lci} typically uses short-term metrics as surrogates for long-term effects~\cite{athey2019surrogate}. Research on \acs{lci} focuses on identifying good surrogates~\cite{mcdonald23,wang22,yang20} or handling confounding factors~\cite{vangoffrier23}.

Among these methods, only Sensitivity Analysis requires an analytic model of the data. Both \acs{iws} and \acs{psm} require access to a sample of the target dataset. \acs{ope} is designed to evaluate new policies. \sat\, however, requires no analytical model and only some global statistics about the target dataset. It is simple, relies on historical data, does not require interventions, does not make assumptions about surrogacy, does not require models of effects, and can be applied to virtually any data, attributes, and metrics.

\section{Experiments}
\label{sec:experiments}

Testing \sat\ poses challenges because it is not feasible to set the
treatment branch of a controlled experiment to match a predefined
scenario, and without treatment results, there is no direct way to
evaluate \sat\ predictions. Therefore, we adopted an indirect
approach: we use an existing experiment and reverse-engineering a
compatible scenario.

In Section~\ref{sec:criteo}, in the interest of reproducibility, we use a publicly available dataset from \criteo. However, important information in the \criteo\ dataset was obfuscated by the company to protect user confidentiality, limiting our ability to analyze the results in full details. Consequently, we also use proprietary data from \spotify\ in Section~\ref{sec:spotify}.

\subsection{Predicting the Probability of Visits on the \criteo\ Dataset}
\label{sec:criteo}

The \texttt{CRITEO-UPLIFT1} dataset~\cite{Diemert2018} was created using data from several incrementality tests, which are randomized trials where a portion of the population is prevented from being targeted by advertising. The dataset contains 25 million rows, each representing a user with twelve features, a treatment indicator, and two binary labels (visits and conversions). Positive labels indicate whether the user visited or converted on the advertiser's website within a two-week test period. The overall treatment ratio is 84.6\%, reflecting advertisers' practice of maintaining a small control population to minimize potential revenue loss. For privacy, the data has been selectively sub-sampled and anonymized to protect the benchmark's competitiveness without revealing the original incrementality level or user context. The dataset is freely accessible on the \criteo\ datasets web page.\footnote{\url{https://ailab.criteo.com/ressources}}

The twelve features, named \cf{0} to \cf{11}, come without descriptions or definitions of how they are computed. We only know that they are predictive of the two labels. In the following experiment, we chose visits as the metric of interest rather than conversions because visits are less rare (4.7\% compared to 0.3\%). This choice helps avoid dealing with extremely skewed data.

In the upper and bottom panes of Figure~\ref{fig:criteo}, we present histograms of the probability of visits in the control and treatment branches. We performed bootstrapping by sampling $B=199$ times with replacement,\footnote{Given that the \criteo\ dataset contains nearly 14 million observations, sampling with or without replacement is unlikely to make a significant difference.} creating  $10,000$ samples from the original dataset and computing the probability of visits for each sample. This procedure provides an estimate of the metric's variability, which is useful for evaluating prediction accuracy.

\subsubsection{The Scenario}
\label{sec:scenario}

To estimate the probability of visits in the treatment branch based on the control data, \sat\ requires a scenario.  This is challenging because \criteo\ did not provide enough details on how the dataset was generated, and building a scenario typically requires domain knowledge. To address this, we identify a scenario likely to produce the control branch data. Essentially, we ask: if we have an ideal or somewhat ideal scenario, how well does \sat\ predict the metrics of interest in the treatment branch?

To keep the experiment simple and representative of a realistic scenario, we impose constraints only on the averages of a subset of features. We identify the features for which the means in the control and treatment branches are significantly different and constrain the weighted averages of this subset to match the corresponding averages in the treatment branch. The results are reported next.

\subsubsection{\sat\ Predictions}
\label{sec:criteo:fortune}

We apply \sat\ to predict the probability of visits in the treatment branch. To establish the constraints, we compute the control set averages and the treatment set averages of the twelve features. Only features~\cf{1}, \cf{4}, \cf{7} and \cf{10} have notably different averages, so we design constraints only for these features.\footnote{We also ran the experiment using constraints based on the twelve features and the results were similar.} Setting these as constraints in Equation~\eqref{form:genalgo} leads to the convex optimization problem in Equation~\eqref{eq:criteo:fortune}, where the sum is over the observations in the control set.

\begin{argmaxi}[2]
  {\bm{\omega}}{\mathcal{H}(\bm{\omega})}
  {\label{eq:criteo:fortune}}{}
  \addConstraint{\frac 1N \sum_n \omega_n \cf{1n}} {=17.00}
  \addConstraint{\frac 1N \sum_n \omega_n \cf{4n}} {=3.59}
  \addConstraint{\frac 1N \sum_n \omega_n \cf{7n}} {=-5.43}
  \addConstraint{\frac 1N \sum_n \omega_n \cf{10n}} {=23.34}
  \addConstraint{\omega_n}{\geq 0}{\quad n \in \{1\dots N\}}
  \addConstraint{\sum_n^N \omega_n}{=1}{\quad n \in \{1\dots N\}}
\end{argmaxi}

Once the weights~$\bm{\omega}$ are evaluated, we use them to estimate the probability of visits~$\tv$ in the treatment set as: \[\etv = \frac 1N \sum_n \omega_n \visit_n\] $\visit_n \in \{0, 1\}$ indicates whether the user in observation~$n$ made a visit, corresponding to the \texttt{visit} column of the \criteo\ dataset.

In practice, it is useful to resample, as we did above, to estimate the probability of visits rather than computing a single point estimate based on the entire control dataset. As described, we sample $B=199$ times $10,000$ observations with replacement from the control set. We then run the procedure that implements~\eqref{eq:criteo:fortune} on each of the 199 data samples. The result is plotted in the pane of Figure~\ref{fig:criteo} titled \sat. While the predicted probability of visits is overestimated, the \sat\ and treatment histograms overlap.

\subsubsection{\ac{nnm}}
\label{sec:criteo:match}

\sat\ only uses the treatment set averages, but suppose instead that
the full treatment branch dataset was available to
us.\footnote{Excluding the key metrics we intend to predict using
  \sat, otherwise the solution would be trivial.} In this hypothetical
case, we should achieve better or equal accuracy compared to using
\sat\ because we have more information about the treatment. This task
is amenable to \acl{nnm} (NNM), also known as greedy matching.

\ac{nnm} is a type of \acf{psm}~\cite{rubin1973matching,hernan2023causal}, a family of methods typically used to identify causal relationships from observational studies. In \acs{psm}, a set of weights is evaluated to match the control as closely as possible to the treatment set. \sat, on the other hand, finds a set of weights that satisfy the scenario, which is designed to approximate the controlled experiment that produced the treatment set. It therefore makes sense to compare the two methods.

The idea behind \ac{nnm} is the following.  For each individual in the treatment group, it finds the most similar individual (or individuals) in the control group  based on a set of observed characteristics — hence the term ``nearest neighbors''. Here, we use the \texttt{matchit} function from the \texttt{MatchIt}~\texttt{R} package~\cite{matchit}, which is based on the propensity score computed using generalized linear model (\texttt{glm}). 

\acs{nnm} cannot be used when we only have a scenario like \sat's. Instead, it requires a full treatment dataset to match the observations. However, we expect it to be more accurate because it utilizes more detailed information, making it an upper bound on what \sat\ can achieve.  

The results of \acs{nnm} are reported in the ``match'' pane of
Figure~\ref{fig:criteo}. We observe that the histogram in the
``match'' pane co\"incides more closely with the ``treatment'' pane
and sits between \sat\ and ``treatment''. Despite this, \sat's
performance is quite good, considering the simple scenario needed to
specify the problem.

\subsubsection{Analysis}
\label{sec:criteo:analisys}

\begin{figure}[th]
  \centering
  \includegraphics[width=.45\textwidth]{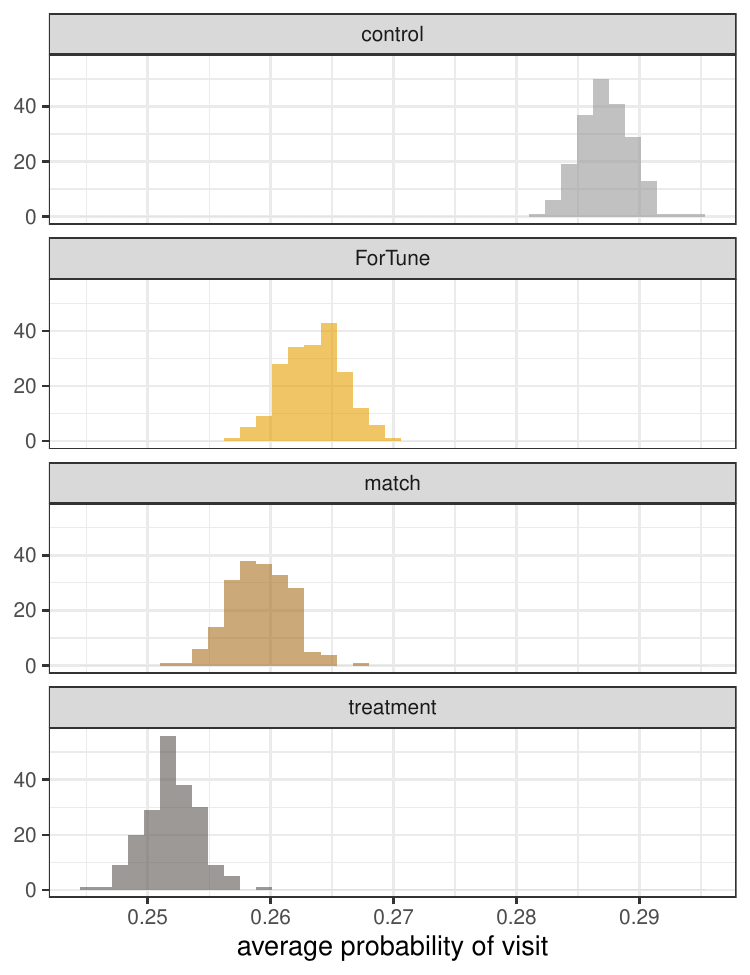}
  \caption{Probabilities of Visit. The ``control'' and ``treatment'' panes report the probability of visit in the control and treatment sets.  The panes titled ``\sat'' and ``match'' show the estimated probability of visit~\etv\ on the treatment set by the respective methods. We observe that even though the ``match'' predictions align better with the histogram in the ``treatment'' pane the ``\sat'' predictions are quite good.}
  \label{fig:criteo}
\end{figure}

The actual probabilities of visits for control and treatment are reported in the top and bottom panes of Figure~\ref{fig:criteo}. Some might be surprised to see that the probability of visits is lower in the treatment group. A possible explanation is that users in the treatment group are targeted less frequently but with more accuracy, leading to both less exposure and higher conversion rates.\footnote{Based on a personal conversation with a former \criteo\ employee.} The predictions from the \match\ and \sat\ methods are plotted in the two middle panes. As expected, \match\ performs better; the histograms generated by the bootstrap runs partially overlap. \sat\ underestimates the effect of the treatment, but the direction is correct and the magnitude is relatively close. 

Overall, considering that the constraints in Equation~\eqref{eq:criteo:fortune} provides only a high level description of the actual treatment effects, \sat's predictions are remarkably close. This result is representative of what we often observe; the predictions are not perfect but point in the right direction, making them useful for informing decision-making.

\subsection{\sat\ at \spotify}
\label{sec:spotify}

\begin{figure}%
  \centering %
  \includegraphics[width=.4\textwidth]{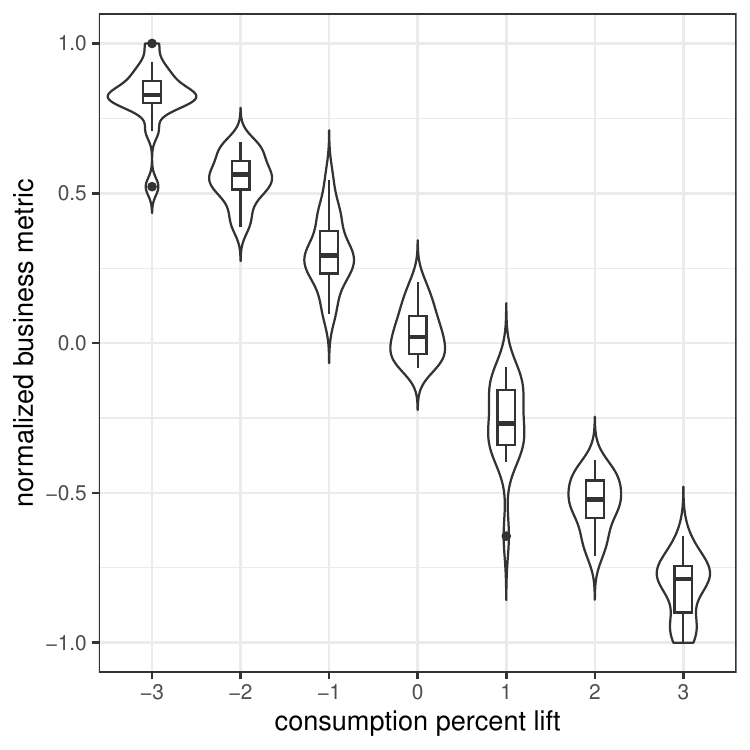}%
  \caption{The business metric is scaled to range between -1 and
    1. Consumption distribution is evaluated by bootstrapping (B=50)
    for each value of the consumption percent lift on the x-axis. The
    distribution is represented both by a violin plot and a regular
    box plot.  The business metric value is distributed around 0 when
    the consumption lift is null. The variability results from
    bootstrapping and gives an estimate of the intrinsic noise in the
    data.}
    \label{fig:spotify:decrease}%
\end{figure}

\spotify\ is among the largest audio streaming services in the world, offering music, podcasts, and audiobooks to users worldwide. As a large business, \spotify\ consists of many different teams, each with its own key performance metrics, some of which may compete with each other. For instance, one team may prioritize driving conversion to the premium product, while another may focus on long-term retention or engagement. These teams run numerous experiments,  typically optimizing for one or two metrics.

Historically, one of the most influential areas within the \spotify\ app is the homepage, where users discover new content and revisit familiar content. Given that changes to the homepage can significantly impact key metrics, we decided to analyze ForTune through a series of experiments conducted on this surface. Specifically, we chose to focus on business metrics, as they are of utmost importance and have historically been less emphasized in favor of easier-to-measure, short-term engagement metrics.

The business estimates reported here are obfuscated to preserve confidentiality. We always apply a monotonic transform to the business estimates and the variables they depend on. Additionally, we only report relative change, not absolute values and normalized them to span the $[-1,1]$ interval. The different figures report results for various user cohorts, but we do not specify how these cohorts are defined. When we report the obfuscated business metric changes against certain feature values, the definitions of these features are intentionally vague. While these measures protect \spotify's business confidentiality, they do not prevent the analysis from illustrating the potential of the method proposed in this paper.

The methodology is the same as in the previous section: we utilize the control branch data and a scenario to predict the metrics of interest observed in the treatment branch. The main result is that out of 10 monitored metrics from 5 experiments run on \spotify's mobile app homepage, 6 were statistically significant. \sat\ estimated mean was directionally aligned 9 times, and the estimated mean histograms overlapped the observed values 8 times. While this is anecdotal evidence, it helped build our confidence in the tool. In Section~\ref{sec:spotify:bad} we revisit the failed experiment to analyze it in more detail.

We proceed by illustrating representative results obtained from these experiment and how they can be used.

\subsubsection{Dealing with Scenario Uncertainty}
\label{sec:spotify:onedimension}

When designing a scenario, it is often challenging to identify precise values for setting constraints. For example, \spotify\ can influence user consumption by altering what is surfaced or by changing the user interface. While we can build some knowledge on the extent of these changes, uncertainty about the exact amplitude of the consumption change always remain.

A natural step then, is to evaluate different scenarios where consumption varies across a range of values that reflects this uncertainty. Figure~\ref{fig:spotify:decrease} shows the relationships between user consumption and a business metric of interest. In this particular case, the relationship is approximately linear and decreasing. It is also important to note that \sat\ is not limited to linear relationships; in other cases, we have observed that the business metric flattens beyond certain values of the control feature.

The information provided by Figure~\ref{fig:spotify:decrease} is useful for deciding how to act on specific cohorts and markets. Depending on whether the slope is positive or negative in a given market, the company might apply different strategies. If the slope is close to zero for business metrics like user retention, user satisfaction, revenue, expenses, this might suggest deprioritizing a project in favor of another one with more promising outcomes.

\subsubsection{Exchange Rates}
\label{sec:spotify:twodimension}

\begin{figure}[t]
  \centering
  \includegraphics[width=.4\textwidth]{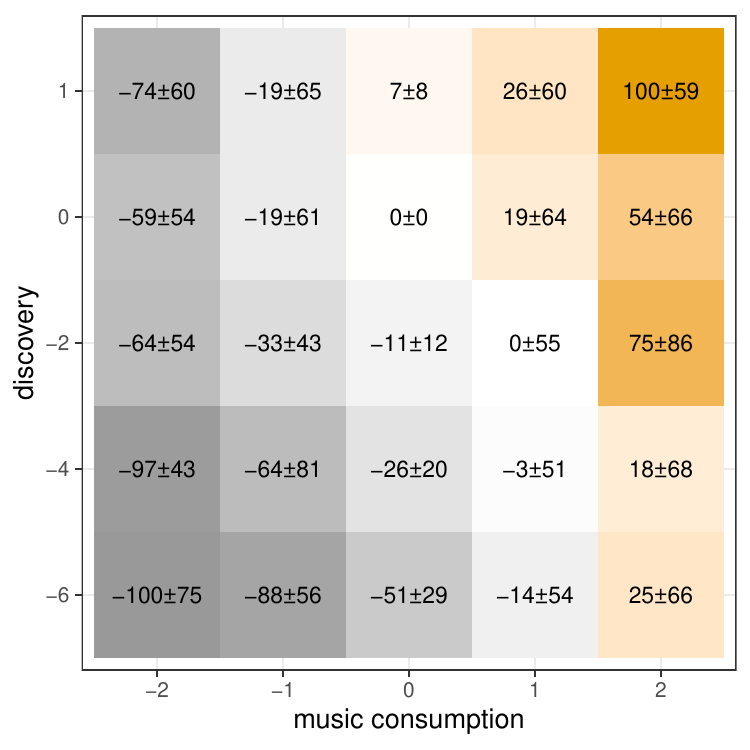}
  \caption[what is a short caption?]{ Scaled User Satisfaction. Estimation of user satisfaction in relation to music consumption and discovery of new content based on 50 bootstraps. }
  \label{fig:spotify:exchange}
\end{figure}

The need to trade off between two metrics naturally arises at many decision points. For example, podcast consumption might compete with music consumption for user time, although increase consumption of one type might expose the user to more opportunities to consume the other type. Predicting which of these opposing effects will dominate is often difficult, especially when the control variables vary in intensity; the trade-offs between podcast and music consumption could differ for users with light versus heavy consumption.

We can use \sat\ to quantify the impact of such trade-offs on business metrics of interest and introduce the concept of an exchange rate between control variables. The objective is to answer the question: for a fixed value of the business metric, how much must podcast consumption change to compensate for a change in music consumption?

Such trade-offs are ubiquitous in complex applications. In web search, advertisements compete with organic search results and the diversity of the result list competes with recall. At \spotify, the discovery of new music competes with users' desire for familiar content, and popular content competes with more personalized niche content. In the example below, we examine the trade-offs between music consumption and the discovery of new content on a metric akin to user satisfaction.

We run a \sat\ experiment with bootstrapping ($B=50$) over a grid of music consumption and discovery values, and report the associated business metric in Figure~\ref{fig:spotify:exchange}. Each cell displays the median of the business metrics and the empirical standard deviation. We observe a clear trend where user satisfaction increases with discovery and with music consumption, even though it is not statistically significant. The figure also suggests that a decrease in the discovery rate below -4 is associated with a more severe decrease in user satisfaction, although this effect is mitigated by an increase in music consumption.

This technique, while still being refined, is used at \spotify\ to
inform decisions about top-level targets for company strategies
related to content discovery and user satisfaction.

\subsubsection{Example of a Miss-Specified Scenario}
\label{sec:spotify:bad}

We examine a case where \sat\ failed to provide a prediction directionally aligned with the treatment branch outcome. We demonstrate how this scenario was improved by adding more constraints, emphasizing the importance of how the scenario is defined, as noted in Section~\ref{sec:algo:limitations}.

In this example, a new ranking function for podcasts was tested on
\spotify's homepage. The experiment showed a negative, statistically
significant impact on two business metrics related to consumption and
subscriptions. The control metrics used in the experiment were podcast
consumption and user activity on the homepage. Based on this scenario,
the tool failed to predict the negative impact on the two business metrics. Instead, it predicted an increase. 

\begin{figure}[t]
  \centering
  \includegraphics[width=.4\textwidth]{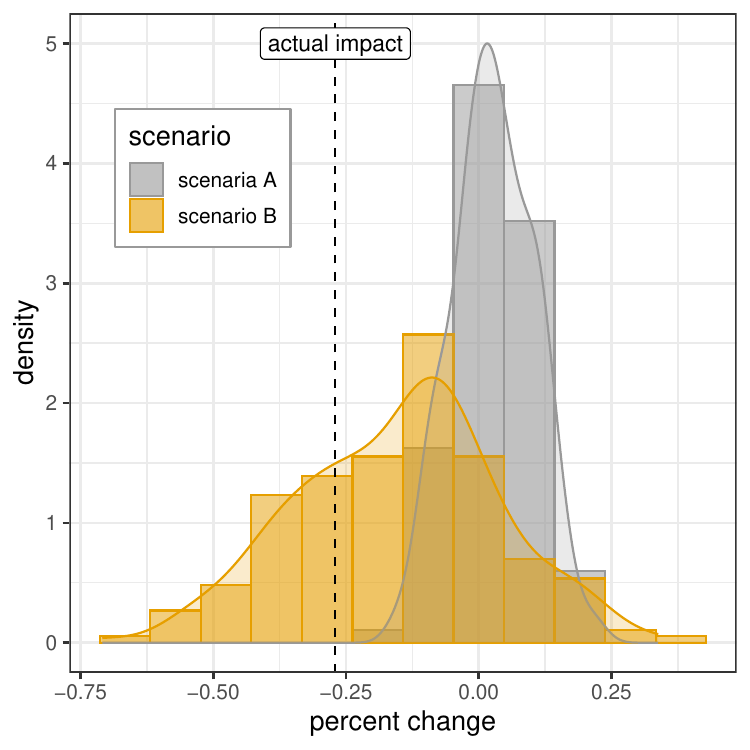}
  \caption{Distribution of the estimations given by two different scenarios and comparison with the true value. Adding more constraints shifted the distribution of predictions for each bootstrap and made the median of the distribution closer to the actual value.}
  \label{fig:spotify:ab_histplot}
\end{figure}

After a thorough analysis, we added constraints to the features related to the consumption of other content types and subscription behavior. This not only led to a more accurate prediction, but also prompted the experiment designers to revise their hypotheses regarding the treatment effect and the specification of control metrics during the experiment. Figure~\ref{fig:spotify:ab_histplot} shows histograms of the predictions before and after adding the extra constraints (Scenarios \texttt{A} and \texttt{B}, respectively). The figure also reports the target metric value observed in the treatment.

\section*{Conclusion}
\label{sec:conclusion}

The limitations of online testing and our desire to produce useful insights for product decisions led us to develop \sat, a flexible, lightweight, and inexpensive approach to investigating hypotheses about changes in consumption behavior, business metrics, and trade-offs between the two. \sat\ is an offline, model-free solution that can explore many hypotheses at low cost, providing powerful support for product leaders making key decisions.

However, like most technological advances, \sat\ comes with trade-offs. It is important to emphasize that not all of \sat's predictions will be accurate or precise, although better domain knowledge and experience help build higher confidence. We expect predictions to have a large variance to accommodate the uncertainty inherent in the vague and under-specified scenarios for which the tool is intended. These scenarios are typically described by a limited number of simple constraints, such as a new average or a new proportion, without addressing the causality of relationships among constraints or user behaviors.  \sat\ is best used to identify trends in trade-offs, extrapolate changes in consumption due to algorithmic changes to longer-term business metrics, and generate hypotheses for deeper analysis.

Nevertheless, teams at \spotify\ are using the tool to generate insights that have previously eluded many teams and leads. These insights have been crucial in making several key product decisions, providing decision-makers with an understanding of relationships between key metrics. We intend to continue expanding the tool by adding new diagnostics, particularly measures derived from information theory, and extending it to end-to-end offline evaluation by linking it to off-policy estimation of consumption shifts and projections of user growth.

\bibliographystyle{ACM-Reference-Format}
\balance%
\bibliography{fortune}

\end{document}